\documentclass[prd,aps,floats,letterpaper,floatfix,superscriptaddress,nofootinbib,twocolumn,preprintnumbers]{revtex4}

\usepackage[mathenv]{}
\usepackage[dvipdf]{graphicx}
\usepackage{amsmath}
\usepackage{color}

\newcommand{\beq}{\begin{equation}}
\newcommand{\bea}{\begin{eqnarray}}
\newcommand{\eeq}{\end{equation}}
\newcommand{\eea}{\end{eqnarray}}

\begin{document}
\title{A squeezed state source using radiation-pressure-induced rigidity}
\author{Thomas Corbitt}
\affiliation{LIGO Laboratory, Massachusetts Institute of
Technology, Cambridge, MA 02139}

\author{Yanbei Chen} \affiliation{Theoretical Astrophysics, California
Institute of Technology, Pasadena, CA 02139}

\author{Farid Khalili} \affiliation{Physics Faculty,
Moscow State University, Moscow 119992, Russia}

\author{David Ottaway}
\affiliation{LIGO Laboratory, Massachusetts Institute of
Technology, Cambridge, MA 02139}

\author{Sergey Vyatchanin} \affiliation{Physics Faculty,
Moscow State University, Moscow 119992, Russia}

\author{Stan Whitcomb} \affiliation{LIGO Laboratory, California
Institute of Technology, Pasadena, CA 02139}

\author{Nergis Mavalvala}
\affiliation{LIGO Laboratory, Massachusetts Institute of
Technology, Cambridge, MA 02139}

\begin{abstract}
We propose an experiment to extract ponderomotive squeezing from
an interferometer with high circulating power and low mass
mirrors. In this interferometer, optical resonances of the arm
cavities are detuned from the laser frequency, creating a
mechanical rigidity that dramatically suppresses displacement
noises. After taking into account imperfection of optical
elements, laser noise, and other technical noise consistent with
existing laser and optical technologies and typical laboratory
environments, we expect the output light from the interferometer
to have measurable squeezing of ~5\,dB, with a frequency-independent squeeze
angle for frequencies below $1$ kHz. This squeeze source is well
suited for injection into a gravitational-wave interferometer,
leading to improved sensitivity from reduction in the quantum
noise. Furthermore, this design provides an experimental test of
quantum-limited radiation pressure effects, which have not
previously been tested.
\end{abstract}

\definecolor{purple}{rgb}{0.6,0,1}
\preprint{\large \color{purple}{LIGO-P030070-00-R}}
\preprint{\large \color{purple}{AEI-2005-088}}

\maketitle

\section{Introduction}
Next-generation gravitational-wave (GW) interferometers, such as
those planned for Advanced LIGO~\cite{pfspie,BC1}, are designed to
have a fifteen-fold improvement in sensitivity over present-day
detectors~\cite{LIGOI}. Advanced detector sensitivity at almost
all frequencies in the detection band is expected to be limited by
{\it quantum} noise~\cite{Caves}. At higher frequencies (above
$\sim 200$\,Hz for Advanced LIGO), quantum noise is dominated by
shot noise, which reflects the accuracy at which test-mass motion
is measured at individual instants; shot noise decreases with
increased input laser power. At lower frequencies (below $\sim
100\,$Hz), quantum noise is dominated by radiation-pressure noise,
which arises from random forces exerted on the test masses by
amplitude fluctuations of the light; radiation-pressure noise
increases with increased laser power. At any given frequency,
spectral density of the quantum noise is a sum of those of the
shot noise, the radiation-pressure noise, and a term arising from
their correlation. The Standard Quantum Limit (SQL) on precise
measurement of the motion arises when the two noise sources are
uncorrelated~\cite{BK92,sql_def}.

Since both types of quantum noise can be attributed to vacuum
fields entering the interferometer from its anti-symmetric
port~\cite{Caves}, injecting squeezed vacuum into this port can
improve the sensitivity of the interferometer~\cite{Caves,Unruh}.
However, for different kinds of interferometers, the required
squeezed vacuum may be very different. For example, for a
Michelson interferometer with Fabry-Perot cavities in each arm
that are tuned to the carrier frequency, and using a homodyne
scheme to detect the phase quadrature of the output light (the
quadrature in which the signal due to differential arm length
changes resides), shot noise is associated with the phase
quadrature of the input vacuum field, while radiation-pressure
noise is associated with the input amplitude quadrature. As a
consequence, a nearly phase-squeezed vacuum is required for higher
frequencies, at which shot noise dominates; while a nearly
amplitude-squeezed vacuum is required for lower frequencies, at
which radiation-pressure noise dominates~\cite{KLMTV}. As another
example, for a narrow-band signal-recycled configuration with
homodyne detection, the squeezed quadrature of the input squeezed
vacuum needs to go through a rapid change from below to above the
optical resonant frequency in order for noise in the detected
output quadrature to be suppressed (instead of amplified)
throughout in this narrow frequency band~\cite{harms}. Moreover,
speed meters have the property that their optimal squeezed
quadrature stays fairly constant for a broad frequency
band~\cite{SM1,SM2,SM3}. Fortunately, it has been shown that
detuned Fabry-Perot cavities can act as optical filters, which
convert a squeezed vacuum with {\it frequency-independent} squeeze
quadrature into one with {\it frequency-dependent} squeeze
quadrature~\cite{KLMTV}, where $\Omega$ is the sideband
frequency~\cite{SM2}. These filters have been shown to be broadly
applicable to existing interferometer
configurations~\cite{KLMTV,harms,SM2,SM3}. Amplitude filters,
which do not rotate the squeeze quadrature, but instead filter out
(i.e., substitute with ordinary vacuum) the squeezed vacuum at
above/below certain frequencies, have also been
analyzed~\cite{ampl_filt}. With these filters as tools, it is
sufficient to construct a device which generates {\em
frequency-independent} squeezed vacuum.

The injection of squeezed light into the antisymmetric port of an
interferometer has been experimentally
demonstrated~\cite{xiao_wu_kimble,mckenzie2002}. In these
experiments, the traditional method for preparing squeezed states
of light using the $\chi^{(2)}$ nonlinearity in optical media was
employed. The squeezed light was generated using optical
parametric processes, and then injected into the antisymmetric
port of the interferometer. The use of detuned Fabry-Perot filters
in generating frequency-dependent squeezed quadratures also has
been demonstrated recently~\cite{chelkowski}. In all of these
experiments, sub-vacuum performance was measured in the few MHz
frequency band, where the deleterious effects of classical noise
sources, such as laser intensity and frequency noise, are greatly
reduced. On the other hand, for GW detection, it is necessary to
inject vacuum states that are squeezed in the GW band, from 10~Hz
to 10~kHz. A recent experimental demonstration of squeezed vacuum
at frequencies as low as 280~Hz~\cite{mckenzie2004} has shown that
low frequency squeezing is possible using optical parametric
processes, but there may be technical limits to the level of
squeezing that can be achieved by this technique, e.g., arising
from photothermally driven fluctuations~\cite{goda2005}.

An alternative technique is to extract the
radiation-pressure-induced -- or {\em ponderomotive} -- squeezing
generated inside an interferometer as a result of the coupling
between the optical field and the mechanical motion of the
mirrors. The properties of the ponderomotive squeezed state depend
on the intensity of the laser light incident on the movable
mirror, optical properties of the interferometer, and on
mechanical properties of the mirror. In this paper, we propose and
analyze a ponderomotive squeezing experiment, which is a variant
of the interferometer that was analyzed in Ref.~\cite{code_paper}.
The main features of this interferometer are high-power optical
field and low-mass mirrors, suspended as pendulums, in order to
enhance the radiation pressure forces; and the use of detuned
Fabry-Perot arm cavities, which induces a optomechanical rigidity,
or {\em optical spring}. Our proposal to use the optical spring
effect is the major innovation over previous attempts to extract
ponderomotive squeezing from
interferometers~\cite{kimble_private}. With our high-power and
low-mass system, the optical spring can be very {\em stiff}, and
will shift the resonant frequency of the test mass from the
suspension pendulum frequency of $\Omega_{\rm p} \sim 2\,\pi
\times 1$~Hz up to $\Theta\sim \,2\,\pi \times 5$~kHz. There are
two main consequences following this shift; for a sideband
frequency $\Omega$ between $\Omega_{\rm p}$ and $\Theta$: (i) all
thermal and seismic forces will now induce much less motion of the
mirror, with a reduction factor of $(\Theta/\Omega)^2$, and (ii)
since the mirrors response to driving forces is frequency
independent at $\Omega < \Theta$, the ponderomotive squeezing
generated in this frequency band is frequency independent.

Experiments with the goal of directly measuring the SQL on the
motion of macroscopic oscillators are similar to the experiment
proposed here in that they must reach a sensitivity that is
limited by quantum-limited radiation pressure. SQL experiments,
however, rely on measuring at a quadrature where the radiation
pressure noise and shot noise remain uncorrelated, whereas this
experiment relies on measuring at a quadrature where the two
noises are correlated. Furthermore, the optical spring in the
ponderomotive squeezing experiment modifies the dynamics of the
system, and allows squeezing to be observed without measuring at
the level of the SQL, which greatly relaxes the sensitivity
requirements compared to the SQL experiments.

The paper is organized as follows: In Sec.~\ref{sect:single_cav}
we discuss the origin of ponderomotive squeezing using a single
Fabry-Perot cavity as a simple but instructive case that explains
many features of our experiment, and will guide our choice of
parameters; in Sec.~\ref{sect:design} we present and motivate the
more complex design of the experiment; in Sec.~\ref{sect:noise} we
calculate contributions from expected noise sources; and in
Sec.~\ref{sect:conclusions} we summarize our conclusions.

\section{Simplified consideration: an optical cavity}
\label{sect:single_cav}

In this Section we consider the ideal case of a short, lossless
Fabry-Perot cavity. For clarity and simplicity, we restrict
ourselves to the quasistatic regime, in which the cavity bandwidth
is much larger than the frequency of observation. This
approximation provides quantitatively correct results in certain
limited test cases.

Consider a Fabry-Perot (FP) cavity with a movable and perfectly
reflective end mirror. Suppose laser light with frequency
$\omega_0$ (the carrier) is incident on a fixed and {\it highly
reflective} input mirror, and assuming the cavity to be {\it close
to resonance,} we list several quantities characterizing the state
of the cavity, namely its linewidth $\gamma$, finesse
$\mathcal{F}$,  circulating power $W$, and the phase shift $\Phi$
gained by the carrier as it comes out from the cavity,  in terms
of more basic parameters:
\begin{eqnarray}
\gamma &=& \frac{c\,{\cal T}_{I}}{4\,L} \,,\\
{\cal F} &=& \frac{2\,\pi}{{\cal T}_I} \,,\\
W(I_0,\delta_\gamma) &=& \frac{4 I_0}{{\cal T}_I}\,\frac{1}{\left(1+\delta_\gamma^2\right)}\,,
\label{W} \\
\Phi(\delta_\gamma) &=& -2\,\arctan \left(\delta_\gamma\right)\,.
\label{Phi}
\end{eqnarray}
Here $L$ is the cavity length,  $\mathcal{T}_I$ the input-mirror
power transmissivity, $I_0$ the incident power, and $c$ the speed
of light. The detuning parameter $\delta_\gamma$,
\begin{equation}
\delta_\gamma \equiv \frac{\delta}{\gamma},
\end{equation}
is defined in terms of $\delta \equiv \omega_{\rm res}-\omega_0$,
the difference between the cavity's (most nearby) resonant
frequency and laser frequency. Note that in Eqs.~\eqref{W} and
\eqref{Phi}, we have explicated the dependence of $W$ on $I_0$ and
$\delta_\gamma$, and the dependence of $\Phi$ on $\delta_\gamma$.
Mathematically, our assumptions of highly reflective input mirror
and the cavity's closeness to resonance amounts to keeping results
up to leading order in $\mathcal{T}_I$ and $\delta L/c$.

The radiation-pressure, or ponderomotive, force $F$ acting on the
end mirror is proportional to the optical power $W$ circulating in
the cavity:
\begin{equation}
  F = \frac{2W}{c} \,.
\end{equation}
For a particular constant set of input power $I_0$ and detuning
parameter $\delta_\gamma$, a DC force acting on the end mirror,
e.g., from the pendulum, can balance the associated ponderomotive
force and keep the mirror in mechanical equilibrium. Now suppose
we shift the mirror {\it statically}, by $dx$, from this
equilibrium condition. Because the detuning parameter
$\delta_\gamma$ will change, the ponderomotive force will also
change, giving rise to an additional restoring force to that from
the pendulum. The total restoring force can be written as (with
$\Omega_{\rm p}$ the pendulum frequency and $M$ the end-mirror
mass):
\begin{equation}
\label{dF}
dF = - M\Omega_{\rm p}^2 \, dx +\underbrace{\frac{2}{c}\frac{\partial W(I_0,\delta_\gamma)}{\partial \delta_\gamma} \frac{d\delta_\gamma}{dx}}_{  \mbox{optical rigidity} \atop {\displaystyle \equiv -K_{\rm opt}}} \, dx\,.
\end{equation}
As we shall see later in this paper, the optical rigidity (or
spring constant) that appears in this equation will be  crucial
for our ponderomotive squeezer.  Note that Eq.~\eqref{dF} is valid
not only for static changes in cavity length, but for all mirror
motions band-limited {\it well below the cavity linewidth} --- the
{\it quasistatic} regime. It is also easy to obtain:
\begin{equation}
\label{dldx}
\frac{d\delta_\gamma}{dx}= -\frac{4\omega_{\rm 0}}{c \mathcal{T}_I}\,.
\end{equation}

We are now ready to set up the frequency-domain equation of motion
for the mirror, at non-zero frequencies well below the cavity
linewidth (i.e., in the {\it quasistatic} regime):
\begin{equation}
\label{mirroreom}
-M\Omega^2 \tilde{x} =
-(M\Omega_{\rm p}^2 + K_{\rm opt}) \tilde {x} +\frac{2}{c}\frac{\partial W(I_0,\delta_\gamma)}{\partial I_0}\tilde{I_0}+\tilde{F}_{\rm ext}\,.
\end{equation}
In this equation, the ponderomotive force associated with AC power
fluctuation $\tilde{I}_0$ and external forces $\tilde{F}_{\rm
ext}$ have been considered. As for the output field, the AC
component of the phase of the output carrier can be written as
[Cf. Eqs.~\eqref{Phi} and \eqref{dldx}]:
\begin{equation}
\label{phieom}
\tilde\Phi  = \left[\frac{d\Phi(\delta_\gamma)}{d\delta_\gamma}\right]\left(\frac{d\delta_\gamma}{dx}\right) \tilde{x}
\end{equation}
Here and henceforth this section, we shall use $(\bar I_0,\bar
\delta_\gamma,\bar\Phi)$ to denote {\it DC components} of the
input power, detuning parameter, and carrier phase shift, and use
$(\tilde{I_0},\tilde \delta_\gamma,\tilde\Phi)$ to denote  their
AC components.

As can already be seen from Eqs.~\eqref{mirroreom} and
\eqref{phieom}, any suspended cavity (not necessarily detuned)
will convert input amplitude fluctuation into mirror motion, and
subsequently output phase fluctuation --- producing ponderomotive
squeezing when the input fluctuations are quantum limited.
Henceforth in this section, we shall further develop and apply
these equations and study main features (in particular advantages)
of a ponderomotive squeezer based on {\it detuned} cavities with
optical rigidity. Before doing that, let us point out that in the
case both mirrors are suspended, we can replace  $M$ in the above
formulas by the reduced mass
\begin{equation}\label{m_red}
  m = \frac{M_{I}\,M_{E}}{M_{I} + M_{E}}
\end{equation}
where $M_{I}$ and $M_{E}$ are the masses of the input and end
mirrors of the cavity, respectively. [Throughout the manuscript we
refer to the input and end mirrors of cavities as input mirror
(IM) and end mirror (EM), respectively. ] We can do so because the
cavity finesse is high, and  the ponderomotive forces acting on
the IM and EM are equal, with a value that only depends on their
relative distance.

\subsection{Input-output relation}

Let us now put the above discussions, in particular
Eqs.~\eqref{mirroreom} and \eqref{phieom}, into the two-photon
formalism~\cite{CS}.  The input field can be written as:
\begin{equation}
  a(t) = (A+a_A)\cos\omega_0 t + a_P\sin\omega_0 t \,,
\end{equation}
where $A$ is the mean amplitude and $a_{A,P}$ are quantum
amplitude and phase fluctuations. It is convenient to normalize a
coherent-state input wave as:
\begin{equation}
\label{vacspectra}
  \hbar\omega_0 A^2 = 2\bar I_0 \,, \quad S_{a_A} = S_{a_P} =  1 \,, \quad S_{a_A a_P}=0\,,
\end{equation}
where $S_{a_A}$, $S_{a_P}$ and $S_{a_A a_P}$ are the single-sided
spectral densities of $a_A$ and $a_P$, and their cross spectral
density, respectively. In the quasistatic regime, the entire
output field $b(t)$ will simply be phase-shifted from $a(t)$ by
$\Phi[\delta_\gamma(t)]$, or
\begin{equation}
  b(t) = (A+a_A)\cos[\omega_0 t- \Phi] + a_P\sin[\omega_0 t-\Phi]\,.
\end{equation}
Decomposing $\Phi$ into its DC ($\bar\Phi$) and AC ($\tilde\Phi$)
components, and treating $\tilde\Phi$ as a small quantity, we
obtain
\begin{equation}
  b(t) = (A+b_A)\cos[\omega_0 t- \bar\Phi] + b_P \sin[\omega_0 t-\bar\Phi]\,,
\end{equation}
with
\begin{eqnarray}
\label{boutraw1}
b_A&=&a_A\,, \nonumber \\
\label{boutraw2}
 b_P &=& a_P +  A\tilde\Phi =
a_P + \left[\frac{4}{\mathcal{T}_I}\frac{1}{1+\bar\delta_\gamma^2}\right]\frac{2A \omega_0 \tilde x}{c}\,,
\end{eqnarray}
where in the second line we have inserted Eq.~\eqref{phieom}.

So far we have essentially put Eq.~\eqref{phieom} into the two photon formalism, let us now further develop Eq.~\eqref{mirroreom}. From Eqs.~\eqref{dF} and \eqref{W}, we have
\begin{equation}
\label{Kopt}
K_{\rm opt} = - \frac{4\omega_0 \bar W}{\gamma Lc}\frac{\bar\delta_\gamma}{1+\bar\delta_\gamma^2}\,.
\end{equation}
From this, we can further define a characteristic frequency,
\begin{eqnarray}
\label{Theta}
\Theta^2 \equiv \frac{K_{\rm opt}}{M} &=&- \frac{1}{\gamma } \frac{4\omega_0 \bar W}{ M Lc}\frac{\bar\delta_\gamma}{1+\bar\delta_\gamma^2} \nonumber \\
&=& -\frac{4\omega_0 I_0 \bar\delta_\gamma }{Mc^2}\left[\frac{4}{\mathcal{T}_I}\frac{1}{1+\bar\delta_\gamma^2}
\right]^2
\,.
\end{eqnarray}
Note that $\Theta$ can either be real ($\bar\delta_\gamma <0$) or
be purely imaginary ($\bar\delta_\gamma > 0$).

On the other hand, the fluctuating part of the power incident on the cavity is
\begin{equation}
  \tilde{I}_0= \hbar\omega_0 A a_A
  \,,
\end{equation}
which induces a fluctuating force of
\begin{equation}
\frac{2}{c}\frac{\partial W(I_0,\delta_\gamma)}{\partial I_0} \tilde I_0
= \left[
\frac{4}{\mathcal{T}_I}\frac{1}{1+\bar \delta_\gamma^2}\right]\frac{2\hbar \omega_0 A}{c}a_A
\label{Fopt}
\end{equation}
on the mirror [Cf.~Eq.~\eqref{mirroreom}].
Inserting Eqs.~\eqref{Kopt}--\eqref{Fopt} into Eq.~\eqref{mirroreom}, we get
\begin{equation}
\label{mirror2p}
M\left[\Theta^2+\Omega_{\rm p}^2-\Omega^2\right]\tilde x
=\left[
\frac{4}{\mathcal{T}_I}\frac{1}{1+\bar \delta_\gamma^2}\right]\frac{2\hbar \omega_0 A}{c}a_A +\tilde{F}_{\rm ext}\,.
\end{equation}
This means the mirror's (complex) mechanical resonant frequencies
will shift from $\pm \Omega_{\rm p}$ to $\pm \sqrt{\Omega_{\rm
p}^2+\Theta^2}$ --- if the latter lie within the quasistatic
regime. Suppose this is true, and that $\Omega_{\rm p}$ is much
lower than  $|\Theta|$, then $\pm \Theta$ gives the mirror
mechanical resonant frequencies. These could correspond either to
a resonance in the usual sense when $\Theta$ is real, or to a pure
instability when $\Theta$ is purely imaginary.

Finally, the  input-output relation of the cavity can be obtained
by inserting  Eq.~\eqref{mirror2p} into Eq.~\eqref{boutraw2}.

\subsection{Quadrature coupling and squeezing}

Assuming no external forces acting on the mirrors, we can put the
frequency-domain input-output relation in a very simple form,
\begin{equation}
\left(
\begin{array}{c}
b_A \\
b_P
\end{array}
\right)
=
\left(
\begin{array}{cc}
1 & \\
-2\mathcal{K}(\Omega) & 1
\end{array}
\right)
\left(
\begin{array}{c}
a_A \\
a_P
\end{array}
\right)
\end{equation}
with a coupling constant
\begin{equation}
\label{KOmega}
\mathcal{K}(\Omega) = \left[\frac{1}{1-(\Omega^2-\Omega_{\rm p}^2)/\Theta^2}\right]\frac{1}{\bar\delta_\gamma}\,.
\end{equation}
Clearly, $\mathcal{K}$ couples the output amplitude and phase
quadratures, and gives rise to squeezing in the output state.

In order to quantify squeezing, we look at the quadrature field
measured by a homodyne detector, which is given by
\begin{equation}
  \overline{2b(t)\cos(\omega_0 t- \bar \Phi - \zeta)}
  = (A+b_A)\cos\zeta + b_P\sin\zeta \,,
\end{equation}
where $\zeta$ is the homodyne angle, with a convention in which
$\zeta=0$ corresponds to the simple amplitude detection of the
output field. The fluctuating part of the output quadrature is
\begin{equation}
  b_A\cos\zeta + b_P\sin\zeta = a_A[\cos\zeta - 2{\cal K}\sin\zeta]
    + a_P\sin\zeta\,,
\end{equation}
with a spectral density of
\begin{equation}
\label{Szeta}
  S_\zeta(\Omega) = 1+2\mathcal{K}^2 - 2 \mathcal{K}[\sin 2\zeta + \mathcal{K}\cos2\zeta] \equiv \xi_{\zeta}^2(\Omega)\,.
\end{equation}
Note that for vacuum state we have $S_\zeta(\Omega)=1$.

By minimizing $\xi_{\zeta}(\Omega)$ over quadratures, we obtain the amplitude squeeze factor
\begin{equation}
\xi_{\rm min}(\Omega) = \frac{1}{|\mathcal{K}(\Omega)| + \sqrt{1+\mathcal{K}^2(\Omega)}}\,,
\end{equation}
which is achieved at
\begin{equation}
\zeta_{\rm min}(\Omega) = \frac{1}{2}\arctan\frac{1}{\mathcal{K}(\Omega)}\,.
\end{equation}

In configurations considered here, the pendulum frequency
$\Omega_{\rm p}$ is always much below $\Omega$ and $|\Theta|$, and
thus negligible. Now we can divide the value of $\Omega$ into
three regimes, if $|\Theta|$ lies within the quasistatic regime
(otherwise only the first regime exists).  {\it First,} when
$\Omega  \ll |\Theta|$, we have a constant  $\mathcal{K}$ of
$1/\bar\delta_\gamma$, which means we have a frequency independent
squeezed state. The amplitude squeeze factor and squeeze angle of
this state are:
\begin{equation}
\label{squeezefactor}
\xi_{\rm min}\left[\Omega \ll |\Theta|\right] =  \frac{|\bar\delta_\gamma|}{1+\sqrt{\bar\delta_\gamma^2+1}}\,,
\end{equation}
\begin{equation}
\zeta_{\rm min}\left[\Omega \ll |\Theta|\right]  = \frac{1}{2}\arctan\bar\delta_\gamma\,.
\end{equation}
{\it Second,} for $\Omega \gg \Theta$, the coupling constant
$\mathcal{K}$ tends to zero and the output state becomes vacuum.
{\it Third}, for $\Omega \sim |\Theta|$, the system goes through a
resonance, with strong squeezing and highly frequency-dependent
squeeze angle, if $\Theta$ is real, and goes through a smooth
transition if $\Theta$ is purely imaginary.\footnote{In reality,
we must also consider the influence from a controller, which is
necessary for stabilizing the detuned cavity, see Sections
\ref{sect:pond_damping} and \ref{sect:controls}.}

Consequently, we obtain a frequency-independent ponderomotively
squeezed source with squeeze factor \eqref{squeezefactor} (which
depends only on the detuning parameter $\bar\delta_\gamma$), and
bandwidth $\Theta$. Although the squeeze factor $\xi_{\rm min}$
can be lowered indefinitely by taking $\bar\delta_\gamma \rightarrow 0
$, the bandwidth $\Theta$ will also drop in this process,
according to Eq.~\eqref{Theta} --- unless input power and/or
cavity finesse are increased.

As discussed in the introduction, such a squeezed state can be
transformed into frequency dependent squeezed states by optical
filters~~\cite{KLMTV,SM2,harms,ampl_filt,chelkowski}. Technically,
the independence in frequency makes it easier to reduce laser
noise, allowing broad-band squeezing,  as we shall discuss in
Section~\ref{sect:lnoise}; it also simplifies our readout scheme.

\subsection{Susceptibility to force noises} \label{sect:optical_spring}

Let us now take into account the influence of noisy external
forces [Cf.~\eqref{mirror2p}]. For the same $\tilde F_{\rm ext}$,
if we denote the mirror's response, {\it in absence of optical
rigidity,} by $\tilde{x}^{(0)}$, then the mirror's response in
presence of optical rigidity can be written as \beq
\label{totallength} \tilde{x} =
-\frac{\Omega^2}{\Theta^2-\Omega^2}\, x^{(0)} \,, \eeq which is
suppressed by a factor $\Omega^2/\Theta^2$, when $\Omega \ll
\Theta$. On the other hand, the transfer function from mirror
motion to output optical field is not modified in any special way
by the optical spring [Cf.~Eq.~\eqref{boutraw2}].  In the end,
optical-field fluctuations caused by external forces on the mirror
at the output port of an optical-spring system is suppressed by
the same factor $\Omega^2/\Theta^2$ from a free-mass system with
comparable circulating power, optical bandwidth, and external
force disturbances.

This dramatic suppression, which applies to seismic noise and all
thermal noises, can easily be as large as two orders of magnitude
in amplitude, and is the most important reason for choosing an
optical spring system as our candidate design for the
ponderomotive squeezer.  Theoretically, such a suppression is
present even when a mechanical spring is used. However, mechanical
springs introduce thermal noise, which are in general orders of
magnitude higher than the vacuum noise associated with optical
springs~\cite{rigidity,BC5}.

\subsection{Radiation-pressure-driven instabilities} \label{sect:pond_damping}

The quasistatic approximation we used in this section cannot
describe the ponderomotive damping associated with optical
rigidity. The sign of this damping is known to be opposite to that
of the rigidity~\cite{BC3}. In case we have a positive rigidity,
the damping will then be negative, leading an oscillatory
instability at the resonance frequency, $\Theta$, with a
characteristic time
\begin{equation}
  \tau_{\rm instab} = \frac{\gamma\left(1+\delta_\gamma^2\right)}{2\,\Theta^2}
    \label{eq:instab}
\end{equation}
It can, therefore, be suppressed by a feedback system acting in
restricted band $\Theta \pm 1/\tau_{\rm instab}$, which is outside
of our frequency band of interest $\Omega \ll \Theta$. The control
system for suppressing this instability is detailed in
Section~\ref{sect:controls}.

High circulating power in the detuned cavities, coupled with high
quality factor ($Q$) mechanical modes of the mirrors, may give
rise another type of radiation-pressure induced
instability~\cite{MSUpla2001}. The motion of the mechanical modes
of the mirror creates phase modulation of the intracavity field,
which are converted into intensity modulation due to the detuning
of the cavity. The intensity fluctuations, in turn, push back
against the mechanical modes of the mirror. This mechanism forms
an optical feedback loop that may become unstable in certain
circumstances. In our case, the most likely form of instability is
that in which the frequency of the mechanical mode is comparable
to the cavity linewidth. This instability, which has been
experimentally observed and characterized for the input mirror
modes of our experiment~\cite{pi_os}, is well outside the
bandwidth of our experiment, and stabilizing it with a narrowband
velocity damping loop should be have little effect on the
experiment. The modes of the end mirror are likely to be too high
in frequency (compared to the cavity linewidth) to become
unstable.

Radiation-pressure-induced torques can also lead to angular
instability. Fabry-Perot cavities with suspended mirrors are
susceptible to a dynamical tilt instability~\cite{sidles_sigg}: as
the cavity mirrors tilt, the beam spots also walk away from the
center of the mirrors, which induces a torque that drives the
mirrors further away. This effect is considered in detail in
Section~\ref{sect:optical_design}.

\subsection{Optical losses}\label{sect:opt_losses}

When a cavity with non-zero losses is considered, the noise
spectrum at the $\zeta$ quadrature becomes
\begin{equation}\label{eq:xi_loss}
S_\zeta^{\rm loss}(\Omega)
  = \frac{{\cal T}_{I}\,S_\zeta(\Omega) + {\cal A}}{{\cal T}_{I}+{\cal A}} \,,
\end{equation}
where $S_\zeta(\Omega)$ is the lossless noise spectrum  of
Eq.~(\ref{Szeta}), and ${\cal A}$ is the total loss per
bounce in the cavity. Assuming that ${\cal A}/{\cal T}_I \ll \xi_{\rm min}$ and ${\cal A} \ll {\cal T}_I$, we have
\begin{eqnarray}
  \xi_{\rm min}^{\rm loss}(\Omega)
  &\approx& \xi_{\rm min}(\Omega) + \frac{\cal A}{2\,{\cal T}_I}.
\end{eqnarray}

\section{Experimental design}
\label{sect:design}

\begin{figure}[t]
\includegraphics[height=8cm]{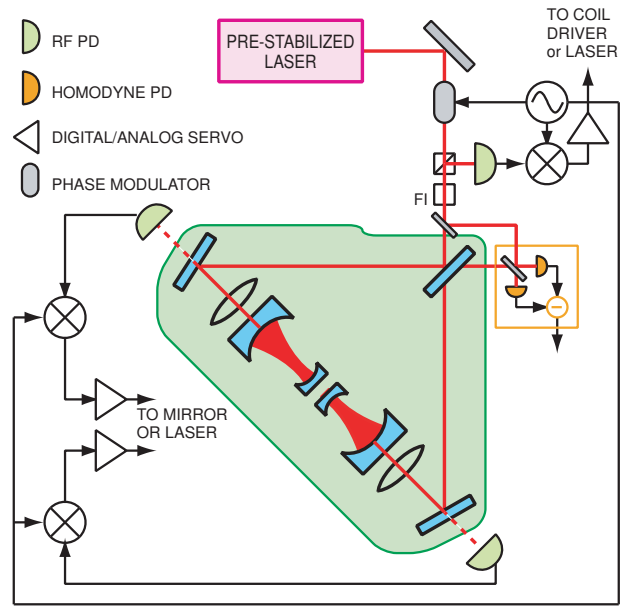}
\caption[Figure] {\label{fig:ponderomotive_layout} Schematic of a
an interferometer designed to extract ponderomotively squeezed
light due to radiation-pressure-induced motion of the low-mass end
mirrors. Light from a highly amplitude- and phase-stabilized laser
source is incident on the beamsplitter. High-finesse Fabry-Perot
cavities in the arms of the Michelson interferometer are used to
build up the carrier field incident on the end mirrors of the
cavity. All interferometer components in the shaded triangle are
mounted on a seismically isolated platform in vacuum. The input
optical path comprises a pre-stabilized 10~Watt laser, equipped
with both an intensity stabilization servo and a frequency
stabilization servo. FI is a Faraday Isolator. }
\end{figure}

In this Section we describe the optical and mechanical design of a
realistic experimental setup for the ponderomotive squeezer. The
interferometer configuration shown in
Fig.~\ref{fig:ponderomotive_layout} is the baseline design for the
experiment. The interferometer is similar to that used in GW
detection: a Michelson interferometer with Fabry-Perot cavities in
each arm. All the mirrors of the interferometer are suspended as
pendulums. While squeezed light could be produced with the use of
a single cavity and suspended mirror, as shown in
Sec.~\ref{sect:single_cav}, the use of interferometry is necessary
to introduce common mode rejection of the laser noise, which would
otherwise mask the squeezed light. Moreover, dark fringe operation
of the Michelson interferometer allows for keeping the dc power
below photodetector saturation levels~\footnote{An alternative
would be to use much lower input power and much higher finesse
cavities, which is generally not feasible.}.

We consider the design features most critical to the goal of
achieving measurable levels of squeezing. The optical design,
described in Section~\ref{sect:optical_design}, includes:

\begin{itemize}

\item A powerful input laser beam with stringent but achievable
requirements on frequency and intensity stability to mitigate
the effects of laser noise coupling;

\item A Michelson interferometer with good contrast for
common-mode rejection of laser noise at the output;

\item Fabry-Perot cavities with

\begin{itemize}

\item high finesse to realize the large optical power incident on
the suspended mirror,

\item substantial detuning (comparable to the cavity linewidth) to create the optical spring,

\item a geometric design that mitigates the effects of
radiation-pressure-induced angular instability;

\end{itemize}

\item An efficient readout chain to detect the
squeezing.

\end{itemize}

The mechanical design of the mirror oscillator, also crucial to
the performance of the interferometer, is described in
Section~\ref{sect:mech_design}.

 \begin{table*}[t]
 \label{table:param}
 \begin{tabular}{llllllll}
 \hline Parameter & \hspace{-0.5cm} Symbol &  Value & Units &
 \hspace{0.2cm} \vline \hspace{0.5cm}  Parameter & \hspace{-0.5cm} Symbol
 &  Value & Units\\
 \hline \hline Light wavelength & $\lambda_{0}$ & $1064$&nm &
 \hspace{0.2cm} \vline \hspace{0.5cm} Input mirror trans. &
 ${\cal T}_{I}$ & $8 \times 10^{-4}$ & --\\
 Input mirror mass & $M_{I}$ & $0.25$&kg & \hspace{0.2cm} \vline
 \hspace{0.5cm} End mirror mass & $M_{E}$ & $1$&g\\
 Arm cavity finesse & $\cal F$ & $8 \times 10^{3}$ & -- &
 \hspace{0.2cm} \vline \hspace{0.5cm}
 Loss per bounce & ${\cal A}/2$ & $5 \times 10^{-6}$ & --\\
 Input power & $I_{0}$ & $4$ & W & \hspace{0.2cm} \vline
 \hspace{0.5cm} Arm cavity detuning & $\delta$ & $1.8\times10^4$&
 rad/sec\\
 BS refl. imbalance & $\Delta_{\mbox{\tiny BS}}$ & $0.01$ & -- &
 \hspace{0.2cm} \vline \hspace{0.5cm}
 Mich. phase imbalance & $\Delta \alpha_{\mbox{\tiny M}}$ & & \\
 Mich. loss imbalance & $\Delta \epsilon_{\mbox{\tiny M}}$ & & &
 \hspace{0.2cm} \vline \hspace{0.5cm}
 Input mirror mismatch & $\Delta_{T}$ & $25 \times 10^{-6}$ & --\\
 Detuning mismatch & $\Delta_{\delta}$ & $10^{-6}$ & $\lambda_{0}$
 & \hspace{0.2cm} \vline \hspace{0.5cm}
 Arm cavity loss mismatch & $\Delta_{\epsilon}$ & $5 \times 10^{-6}$ & --\\
Laser intensity noise & -- & $10^{-8}$& Hz$^{-1/2}$ &
 \hspace{0.2cm} \vline \hspace{0.5cm} Laser phase noise & -- &$10^{-6}$ &--\\
\hline Susp. resonant freq. & $\Omega_{0}$ & $0.7$& Hz &
 \hspace{0.2cm} \vline \hspace{0.5cm} Susp. mech. Q & $Q$ &
 $10^{5}$ & --\\
Parallel coating loss angle & $\phi_{\parallel}$ &
$4\times10^{-4}$&  &
 \hspace{0.2cm} \vline \hspace{0.5cm} Perpendicular coating loss angle & $\phi_{\perp}$ & $4\times10^{-4}$ & \\
Substrate Young's modulus & $Y$ & $7.3\times10^{10}$& N\,m$^{-2}$
&
 \hspace{0.2cm} \vline \hspace{0.5cm} Coating Young's modulus & $Y^{\prime}$ &$1.1\times10^{11}$ & N\,m$^{-2}$\\
Coating thickness & $d$ & $10$& $\mu m$ &
 \hspace{0.2cm} \vline \hspace{0.5cm} Beam radius & $w$ &$1$ & mm\\
Detection loss &$\epsilon_{det}$  & 0.1 & &
 \hspace{0.2cm} \vline \hspace{0.5cm} Temperature & $T$ &$293$ & K\\
 \hline
 \end{tabular}
 \caption{Select interferometer parameters and the nominal values
 we assume for them.}\label{table:parameters}
 \end{table*}

\subsection{Optical design}
\label{sect:optical_design}

The optical configuration is shown in
Fig.~\ref{fig:ponderomotive_layout}, and upper section of
Table~\ref{table:parameters} lists the optical parameters that we
assume in designing the experiment.

\subsubsection{Detuned arm cavities}

The optical spring is the predominant feature of the detuned arm
cavity --- which has been analyzed in detail in
Sec.~\ref{sect:single_cav}. In particular, when a cavity is
detuned, the optical spring modifies the response function of the
differential mode from a free mass (here we ignore the pendulum
frequency) to a harmonic oscillator with resonant frequency
$\Theta$ [see Eq.~(\ref{Theta})].
Our frequency band of interest is $\Omega \ll \Theta$, in which the
response of cavity lengths to external disturbances (e.g., driven
by seismic and/or thermal forces) is suppressed by
$\Theta^2/\Omega^2$, and the (ideal) output state is a
frequency-independent squeezed vacuum with squeeze factor as a
function of $\bar\delta_\gamma = \delta/\gamma$ [Eq.~(\ref{squeezefactor})]. Based on this
qualitative understanding, in order to obtain a substantial
squeeze factor up to around 1\,kHz, we need to choose an optical
configuration such that $\Theta$ is at least several kHz, and
$\delta$ of the same order of magnitude as $\gamma$. This lead us
to a high-power, low-mass, substantially detuned arm cavity.

We have chosen to realize our optical-spring squeezer by a
Michelson interferometer with Fabry-Perot cavities formed by a
large, suspended mirror as the input mirror (IM), and a small,
light, highly reflective mirror as the end mirror (EM). The EM is
chosen to be $1$\,g, as light as we deem possible with current
experimental techniques. We note that the optical spring could
also be created with a detuned signal recycling mirror, as is done
in Advanced LIGO~\cite{BC1}, but that would require an additional
mirror and optical cavity, increasing the complexity of the
system. The suspensions are primarily necessary to allow the
mirrors to behave as free masses in the experimental frequency
band, but also have the added benefit of isolation from seismic
noise. To achieve these benefits, a pendulum resonant frequency of
$0.7$ Hz is chosen. The arm cavities must be placed in vacuum
chambers due to the high finesse and circulating power, and also
to meet the length stability requirements. The mechanical design
of the suspension of the end mirror is discussed in the next
section.

Next we discuss the optical parameters of system. We first set an
``ideal'' target squeeze factor of 17\,dB, i.e., the squeeze
factor of the system in absence of optical losses and technical
noises. This allows for the contribution of the vacuum
fluctuations from the anti-symmetric port to the total noise to be
small. This determines $\delta_\gamma \approx 0.31$. As a next
step, we fix the finesse of the arm cavity, which should be high
because we would like to have the optical-spring resonance
$\Theta$ as high as possible, for a better noise suppression.
Although this could be achieved by increasing input power alone,
it is much more efficient to increase the finesse, because $\Theta
\propto \sqrt{I_0}/{\cal T}$, see Eq.~(\ref{Theta}), note that we
need to maintain $\bar\delta_\gamma$ for a fixed target squeeze
factor); a higher {\em input power} is also undesirable because of
the associated increase in amplitude and phase noise. On the other
hand, cavities with too high a finesse will limit the output
squeeze factor through increased optical losses, and will also
increase the instability from the optical spring. In the end, we
set the transmission of the input mirror to be $800$\,ppm, which,
if assumed to be the dominant loss in the cavity, gives a finesse
of $8\times 10^3$. In this system, for a 4\,W input laser power,
we have a circulating power of roughly 9\,kW, and $\Theta \approx
2\,\pi \times 5$~kHz.

\subsubsection{Angular instability} \label{sect:ang}

Our discussion of the optical properties of the cavities so far
has been restricted to the longitudinal resonances. In this
section we consider the geometrical properties of the cavity,
necessary to avoid angular instability due to
radiation-pressure-induced torque~\cite{sidles_sigg}. For a cavity
with two spherical mirrors, the equations of motion of the two
mirrors are rather straightforward, if the motion frequency is
much lower than the cavity bandwidth (which is trivially true in
our case). Suppose $\theta_{I,E}$ are the tilt angles of two
mirrors with radii of curvature $R_{I,E}$, separated by $L$, then
the equations of motion of $\theta_{I,E}$ are given by (here and
henceforth we denote IM by $I$ and EM by $E$)
\beq \left(\begin{array}{c}\ddot\theta_I \\
\ddot\theta_E\end{array}\right) =\mathbf{M}\left(
\begin{array}{c}
\theta_I \\
\theta_E
\end{array}
\right)\,, \eeq with \beq \mathbf{M}=\frac{1}{1-g_I g_E}\left(
\begin{array}{cc}
g_E \omega_I^2  & -\omega_I^2 \\
-\omega_E^2 & g_I \omega_E^2
\end{array}
\right) - \left(
\begin{array}{cc}
\omega_I^2 & \\
& \omega_E^2
\end{array}
\right)\,. \eeq Here $\Omega_{I,E}$ are the resonant frequencies
of the tilt degrees of freedom of the mirrors in the absence of
radiation pressure,\footnote{We consider two types of tilt angles,
{\it pitch} and {\it yaw}, described in
Section~\ref{sect:mech_design} for our mirrors. In the ideal
situation, pitch and yaw are orthogonal degrees of freedom and can
be considered separately. The resonant frequencies of the IMs and
EMs when they are ``free'' masses, $\Omega_{I,E}$, will, however,
differ from each other, as will the pitch and yaw mode frequencies
for each optic.} and $g_{I,E}$ are the $g$-factors, defined by
\beq g_k = 1-\frac{L}{R_k}\,,\qquad k=I,E. \eeq The angular
frequencies $\omega_{I,E}$ are given by \beq \omega_k^2 \equiv
\frac{2 I_c L}{cJ_k}\,,\quad k=I,E\,, \eeq where $J_k$ are the
moments of inertia of each mirror along the tilt axis under
consideration. These frequencies set the time scales of
tilt-induced dynamics associated with each mirror. In
Table~\ref{table:instab}, we list the relevant parameters for our
IM and EM, along with the resulting $\omega_k$. Note that
$\omega_{\rm E}$ does seem to be in a regime (a few Hz) where we
have to worry about tile instability.  As pointed out by Sidles
and Sigg~\cite{sidles_sigg}, in the absence of external restoring
forces, (i.e., as $\Omega_{I,E} \rightarrow 0$), we have \beq \det
\mathbf{M} = - \omega_I^2 \omega_E^2/(1-g_I g_E) < 0\,, \eeq which
means $\mathbf{M}$ always has one positive eigenvalue (pure
instability) and one negative eigenvalue (stable resonant mode).
On the other hand, the $\Omega_{I,E}$ terms, if large enough, will
stabilize the system.

\begin{table}
\begin{tabular}{c|c|c|c|c|c}
k& $r$ (cm) & $d$ (cm) & $M_k$ (g) & $J_k$ (g$\cdot{\rm cm}^2$) &{$\omega_k/(2\pi)$ (Hz)} \\
&                  &    &       &          & $W=9\,$kW\\
\hline
IM & 4.25 & 2.00 & 250 & 1211 & $0.11$ \\
EM & 0.60 & 0.30 & 1.00 & 0.098 &  $12.4$
\end{tabular}
\caption{Moments of inertia of the mirrors along their tilt axes.
we model the mirrors as a cylinder with radius $r$ and thickness
$T$, and $J=M r^2/4 + M T^2/12$. Circulating powers of 9~kW
 are assumed. \label{table:instab}}
\end{table}

Let us first examine the case without external restoring force.
The resonant frequencies are in general given by

\begin{eqnarray}
\omega_{\pm}^2 &=& \frac{1}{2(1-g_I g_E)}\bigg[-(g_E\omega_I^2+g_I\omega_E^2) \nonumber \\
&&\pm \sqrt{(g_E\omega_I^2+g_I\omega_E^2)^2+4\,(1-g_I
g_E)\omega_I^2\omega_E^2} \bigg]\,.\;
\end{eqnarray}

Noticing that we have $\omega_{\rm I}^2/\omega_{\rm E}^2 \approx
8\times 10^{-5} \ll 1$, we can expand the unstable resonant
frequency up to the leading order in $\omega_{\rm I}^2/\omega_{\rm
E}^2$. We also have to require that $g_{\rm E}$ is not very close
to $0$ ($|g_{\rm E}| \gg \omega_{\rm I}^2/\omega_{\rm E}^2$). Now
if we pay attention only to $\omega^2_{-}$, which is the unstable
resonant frequency, then we have \beq \omega^2_{-} = \left\{
\begin{array}{cc}
\displaystyle -\frac{g_{\rm I} \omega_{\rm E}^2}{1-g_{\rm I}g_{\rm
E}} & g_{\rm I},g_{\rm E}>0
\\
\\
\displaystyle \frac{\omega^2_{\rm I}}{g_{\rm E}} &  g_{\rm
I},g_{\rm E}<0\,.
\end{array}
\right. \eeq This confirms, in our special case, that cavities
with negative $g$ factors are less unstable, as argued by Sidles
and Sigg~\cite{sidles_sigg}. Moreover, each mirror itself, when
the other mirror is held fixed, is stable in the case of negative
$g$-factors (since diagonal elements in $\mathbf{M}$ are both
negative).

Now let us study the stability when external restoring forces are
available. In general the resonant frequencies $\omega$ are given
by \beq \det{\left[\mathbf{M}+\omega^2\mathbf{I}\right]}=0\,. \eeq
The stability condition can be stated more formally as having
$\mathbf{M}$ negative definite, which means requiring
\bea
\label{stab1}
(1-g_I g_E)\omega_I^2 -g_E\omega_I^2 &>& 0 \\
\label{stab2}
(1-g_I g_E)\omega_E^2 -g_I\omega_E^2 &>& 0\\
\det \mathbf{M} &>&0\,, \eea with

\bea
&&\det \mathbf{M}>0\nonumber\\
&\Leftrightarrow &
\left[\omega_I^2-\frac{g_E\omega_I^2}{1-g_Ig_E}\right]
\left[\omega_E^2-\frac{g_I\omega_E^2}{1-g_Ig_E}\right]
>\frac{\omega_I^2\omega_E^2}{(1-g_Ig_E)^2}\,.\nonumber\\
\label{detMcond} \eea

For negative $g$-factor cavities, which start out to be less
unstable, the stabilization is easy: Eqs.~\eqref{stab1} and
\eqref{stab2} are automatically satisfied (since the diagonal
elements are already negative in absence of external restoring
force),  while Eq.~\eqref{detMcond} can be satisfied without
requiring any EM external stabilization, if \beq \label{eq:stable}
\Omega_{\rm I} > \omega_{\rm I}/|g_{\rm I}|\,, \eeq Stabilization
is less straightforward for positive $g$-factor cavities:
$\Omega_{I,E}$ will have to be at least of the same order as
$\omega_{I,E}$, unless we fine-tune $g_{I,E}$. For example,
Eqs.~\eqref{stab1} and \eqref{stab2} already impose \beq
\Omega_{I,E} > \sqrt{\frac{g_{2,1}}{1-g_{1}g_E}}\omega_{I,E}\,,
\eeq which suggests that $\Omega_{\rm E}$ will have to be at least
comparable to $\omega_{\rm E}$, unless we make $g_{\rm I}$ very
small, which is undesirable due to decreased stability of spatial
optical modes. Defining \beq \Omega_{I,E}^2 = (1+\sigma_{I,E})
\frac{g_{E,I}\,\omega_{I,E}^2}{1-g_I g_E}\,,\quad(g_{I,E}>0)\,,
\eeq the stability condition can be written as \beq \sigma_{\rm
I}>0\,, \; \sigma_{\rm E} >0\,,\; \sigma_{\rm I} \sigma_{\rm E} >
\frac{1}{g_I g_E}\,. \eeq

For stability reasons, we propose using negative $g$-factor
cavities. To minimize the angular instability and simultaneously
maximize the beam spot size at the mirrors in order to reduce the
effects of the coating thermal noise, as discussed in
Section~\ref{sect:int_coat_noise}, we propose cavities of length
$L \sim 1$~m, with the mirrors having a radius of curvature
slightly greater than $0.5$ m, in order to have $g \sim -0.8$.

From Eq.~(\ref{eq:stable}), we find a stabilizing IM frequency of
$0.12$ Hz, which is trivially satisfied, to be
sufficient to stabilize the system without an active control
system.

\subsubsection{Optical readout}
\label{sec:readout}

Ideally, the squeezed field would be measured at the antisymmetric
port with a homodyne detector. In this setup, a strong local
oscillator (LO) field is mixed on a beamsplitter with the squeezed
field, and the two resulting fields are measured by photodiodes
and the resulting photocurrents are subtracted, eliminating the
component of the signal due to the LO alone. This scheme is
advantageous because it allows for an arbitrary quadrature of the
squeezed field to be measured, simply by changing the phase of the
LO. The disadvantage of this scheme, however, is that the LO field
must be much stronger than the carrier component of the squeezed
field. Due to mismatches in the system, a portion of the carrier
light will couple out the antisymmetric port. With the parameters
for contrast defect and other optical imperfection listed in
Table~\ref{table:parameters}, we expect the carrier light at the
output to be on the order of $1 {\rm ~mW}$. While a LO level that
is an order of magnitude larger is readily achievable, we begin to
reach the saturation limits of our photodetectors.

An alternative readout scheme is to simply measure the squeezed
field with a photodetector. In this scheme, only the amplitude
fluctuations of the light exiting the antisymmetric port may be
measured. However, our optimization scheme for laser noise, as
described in Sec \ref{sect:lnoise}, has the side effect of
aligning the squeezed quadrature with the amplitude quadrature of
the light exiting the antisymmetric port. While this limits us to
measuring only the amplitude fluctuations of the light, this is
precisely the quadrature in which the squeezing occurs. The
homodyne readout scheme is preferable, but the direct readout is a
viable alternative to avoid power constraints.

In practice, since we wish to control the interferometer degrees
of freedom, we use the detection scheme shown in
Fig.~\ref{fig:feedback_loop}. A small fraction of the
antisymmetric port light ($R \ll 1$ in power) is sampled to
generate an error signal for the control loop, while the majority
is preserved for injection into an interferometer or for detection
of squeezing using either the homodyne or direct detection methods
described above.

\subsection{Mechanical design}
\label{sect:mech_design}

\begin{figure}[t]
\includegraphics[height=6cm]{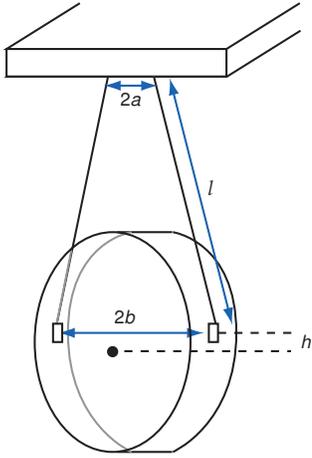}
\caption[Figure] {\label{fig:suspension} Front and side view of
the end mirror suspension. The dot represents the center of mass
of the mirror. The fibers are attached to a point a distance $h$,
which could be negative, above the mirror center of mass. The
distance between the attachment points at the mirror is $2b$, and
at the top of the suspension is $2a$. Not drawn to scale.}
\end{figure}

Both the input and end mirrors of the cavities are suspended from
pendulums. The input mirrors have a mass of 250~g and a 75~mm
diameter; they are identical to the suspended optics used in the
input modecleaner of the initial LIGO
detectors~\cite{s1_detector}. Greater care must be taken in the
suspension of the end mirrors of the cavities, however -- due to
their small mass of $1$~g, the EMs have greater susceptibility to
thermal noise. We use a monolithic fused silica suspension, in
which thin fused silica fibers are welded to the side of the
mirror substrate using a ${\rm CO}_2$ laser. This technique has
been demonstrated to produce a pendulum mode Q of approximately
$10^{7}$~\cite{freytsis}. The suspension design consists of two
fibers, each approximately $10 \,\mu{\rm m}$ in diameter, welded
or glued to the mirror, as shown in Fig.~\ref{fig:suspension}.

To maintain high circulating power in the arm cavities, and
minimize interference from higher-order spatial modes, alignment
of the mirror is critical. Controlling the pitch (rotation about
the horizontal diameter of the mirror) is a particularly important
consideration, since we expect large pitch angles due to static
displacement of the EM with 9~kW of laser power
impinging on it. The frequency of the pitch mode is determined by
the location of the attachment point between the fiber and the
mirror substrate, and the diameter of the fiber~\cite{gonzalez}.
For our regime of fiber lengths, typically 0.5~m, the frequency of
the pitch mode frequency, is approximately
 \begin{equation}
\omega_{pitch} = \sqrt{\frac{T\left(h + \Delta\right)}
{J_{pitch}}},
\end{equation}
assuming $\Delta+h \ll l$,  where $\Delta$ is the characteristic
length at which the fiber bends above its attachment point, $h$ is
the distance of the attachment point from the mirror center of
mass, $l$ is the length of the suspension wire, $T$ is the tension
in the fiber, and $J_{pitch}$ is the moment of inertia for the
pitch degree of freedom (given in Table~\ref{table:instab}). A
higher frequency, $\omega_{pitch}$, will require a larger force to
control the pitch of the mirror. Minimizing the necessary force,
and hence $\Delta+h$ is desirable to limit the actuator range. For
fibers with a diameter of $100 \,\mu {\rm m}$, $\Delta \approx
8.5$~mm, while for $10 \,\mu {\rm m}$, $\Delta \approx 8.5\times
10^{-2}$~mm. In the $100 \,\mu {\rm m}$ case, it would be
impossible to make $\Delta+h$ smaller than a few millimeters,
while for the $10 \,\mu {\rm m}$ case, it can be made very small
by choosing $h$ appropriately. Consideration of the necessary
torques that must be supplied, and the torques that may be
generated by actuators, as well as the ability to create and work
with thin fibers, leads to a choice of fiber diameter of
approximately $10 \,\mu {\rm m}$. Taking $\Delta+h = 100\,\mu {\rm
m}$, $J_{pitch} = 0.098~{\rm g\,cm}^2$, $T= 98$~dyne, we get
$\omega_{pitch} \approx 2 \, \pi \times 0.50$ Hz.
The yaw frequency, again assuming that $\Delta+h \ll l$, is
\begin{equation}
\omega_{yaw} = \sqrt{\frac{2\, T\, a\, b}{l\,J_{yaw}}}
\end{equation}
where $2a$ is the separation between attachment points of the
fibers at the top end of the suspension, $2b$ is the distance
between the attachment points on either side of the mirror, and
$J_{yaw}$ is the moment of inertia for the yaw degree of freedom.
For $a = 6~\rm{mm}$, $b = 3 ~\rm{mm}$, $J_{yaw} = 0.098
~\rm{dyne}$, we get $\omega_{yaw} \approx 2\, \pi \times 0.43$ Hz.

Control of the longitudinal motion of the end mirror is a
difficult task. When the $9$~kW of power in the
cavity is incident on the end mirror, the mirror feels a constant
force, which must be balanced. We choose to balance the constant
(dc) radiation pressure force with gravity. When the mirror is
displaced by a few millimeters from its equilibrium (with no laser
light present), for a given (fixed) pendulum length, the
gravitational restoring force will be equal to the constant
radiation pressure force. In order to lock the cavity at full
power, we propose the following scheme: First, we use an
electromagnetic actuator to offset the mirror the required
distance from its equilibrium position. Next, we lock the cavities
with very small circulating powers, such that the radiation
pressure forces are negligible. We slowly increase the power in
the system, which increases the radiation pressure forces on the
mirrors. Simultaneously, we reduce the pulling force of the
actuator, which will be counteracted by the increasing radiation
pressure force, keeping the mirror at a fixed position. When the
power reaches its design value, the mirror is held in place by a
balance of the radiation pressure, gravitational restoring, and
electromagnetic forces. This provides a way of controlling the
longitudinal degree of freedom of the mirror.

\section{Noise Couplings}
\label{sect:noise} 
\begin{figure*}[t]
\includegraphics[width=1\textwidth]{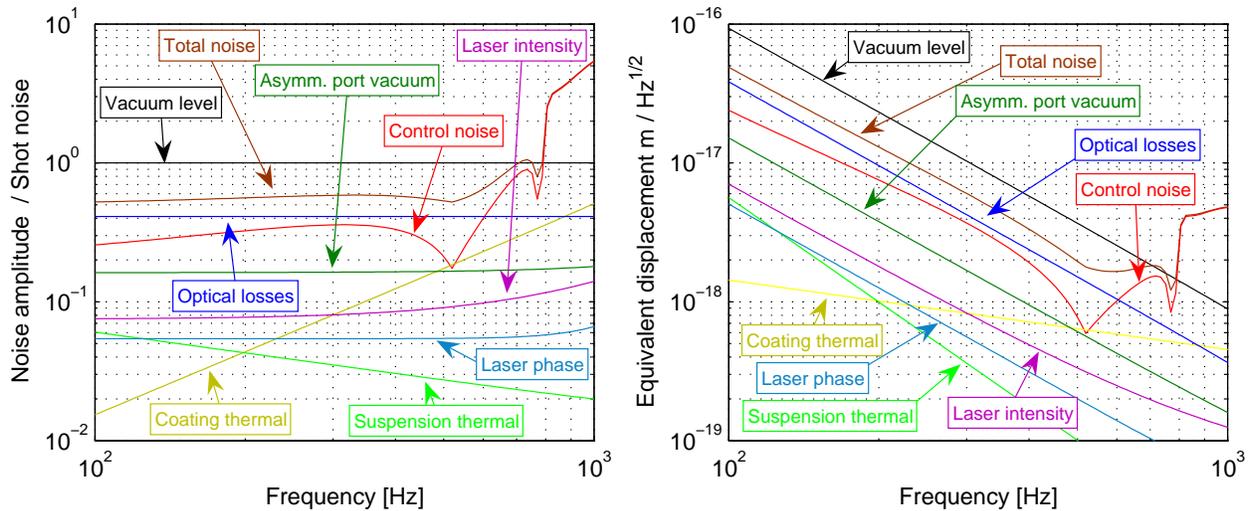}
\caption[Figure] {\label{fig:fig1} Left panel: The different noise
sources relative to the vacuum level, as a function of frequency.
The dominant noise below 1 kHz is optical losses, which are
primarily comprised of detection losses (10\%) and the
optimization losses (13\% in one arm). Right panel: The same noise
sources in terms of equivalent displacement of a free mass (the
displacement noise that each noise source would exhibit if the
optical spring were not present). We estimate that a sensitivity
of $5 \times10^{-16}\mathrm{m\,Hz^{-1/2}}$ is necessary to measure
squeezing at 100~Hz, and the required sensitivity drops as
frequency to the second power at higher frequencies.}
\end{figure*}

In this section, we estimate the contribution
of expected noise sources to the total noise budget. These include
thermal noise from the suspended mirrors (including thermal noise
from the optical coatings on the substrates), as well as laser
intensity and phase noise. In Fig.~\ref{fig:fig1}, we show the
spectral density of the dominant noise sources both in terms of
noise power relative to the vacuum level in a given quadrature,
and also in terms of (free mass) displacement, which does not
include the suppression from the optical spring. Furthermore, we
shall see that the coupling of laser noise has a very strong
dependence on the quadrature to be measured. Careful choice of the
measurement quadrature is critical to successful extraction of the
squeezing; this is analyzed in Section~\ref{sect:lnoise}.

\subsection{Suspension thermal noise}
Applying the Fluctuation Dissipation Theorem \cite{fdiss} to an
object of mass $M$ that is suspended from a pendulum with
mechanical quality factor $Q$ and resonant frequency $\Omega_R$,
we get the free mass displacement noise spectrum~\cite{Saulson}
\begin{equation}
S_{susp}\left(\Omega\right) =
\frac{4\,k_{B}\,T\,}{M\,\Omega\,Q}\frac{\Omega_{R}^2}{\left(\Omega_{R}^{2}-
\Omega^{2}\right)^{2}+\frac{\Omega_{R}^{4}}{Q^2}},
\end{equation}
where $T$ is the temperature and $k_B$ is the Boltzmann constant.
The monolithic fused silica suspension, described in
Section~\ref{sect:mech_design}, is used primarily to reduce
$\phi$. Metal wires and alternative methods of attachment have
higher losses, which would make the suspension thermal noise more
severe. As shown in the curve labelled ``Suspension thermal'' in
Fig.~\ref{fig:fig1}, the monolithic fused silica suspension will
place the suspension thermal noise at a level where it does not
have any measurable effect on the experiment.

\subsection{Internal and coating thermal noise}
\label{sect:int_coat_noise}

The free mass displacement noise spectrum due to internal and
coating thermal noise has been approximated as~\cite{coatingnoise}
\begin{eqnarray}
S_{ICTN}\left(f\right) = & \displaystyle \frac{2k_{B}T}
{\pi^{3/2}f}\frac{1}{w\,Y}\bigg[\phi_{substrate}\nonumber\\
+ & \displaystyle
\frac{d}{w\sqrt{\pi}}\left(\frac{Y^{\prime}}{Y}\phi_{\parallel} +
\frac{Y}{Y^{\prime}}\phi_{\perp}\right) \bigg].
\end{eqnarray}

We assume that $\phi_{substrate} \ll
\frac{d}{w\sqrt{\pi}}\left(\frac{Y^{\prime}}{Y} \phi_{\parallel} +
\frac{Y}{Y^{\prime}}\phi_{\perp}\right)$, so that the dominant
thermal noise is due to the optical coating. Using the parameters
listed in Table~\ref{table:parameters}, we calculate the coating
thermal noise shown in Fig.~\ref{fig:fig1}. We note that the
coating thermal noise is potentially a limiting noise source near
$1$~kHz.

\subsection{Control System Noise} \label{sect:controls}

As discussed in Section~\ref{sect:pond_damping}, the
optomechanical resonance is unstable, i.e., it grows in time, with
typical time scale for instability given by Eq. (\ref{eq:instab}).
This instability must be controlled by use of a feedback loop that
stabilizes the unstable resonance by a damping-like control force.

Defining $s= j\,\Omega$, the transfer function $P(s)$ of the
pendulum, including the optical spring effect, is given by
\begin{equation} \label{eq:pend}
P(s) = \left[s^{2} + \frac{\Theta^{2}
\gamma^{2}}{\left(\gamma+s\right)^2+\delta^2}\right]^{-1}.
\end{equation}

This transfer function is straightforward to interpret; it is the
transfer function of an ideal spring, with a spring constant that
is filtered by the cavity pole. In the limiting case that $\gamma
\gg s$ and $\gamma \gg \delta$, the transfer function of an ideal
pendulum is obtained. This transfer is unstable because it has
poles in the right half plane (the real part of the pole is
greater than $0$).

\begin{figure}[t]
\includegraphics[width=8cm]{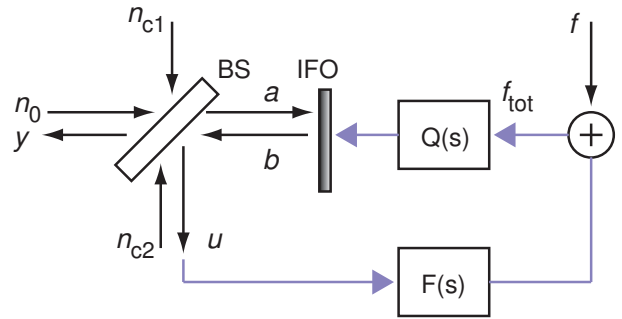}
\caption[Figure] {\label{fig:feedback_loop} Block diagram for the
feedback loop. $a$ and $b$ are the input and output quadrature
fields; $n_{i}$ are vacuum noise fields entering the different
port of the beamsplitter (BS) that has power reflectivity $R$ and
transmission $T$. A small fraction of the output (squeezed) field
$ u = \sqrt{R}\,b$ is used to generate a shot-noise-limited error
signal for a feedback loop to control the position of the
differential mode of the ponderomotive interferometer (IFO), while
the remainder $y = \sqrt{T}\,b$ is used to make
sub-quantum-noise-limited measurements. The sample beam $u$ is
filtered by $F(s)$, a transfer function that converts quadrature
fields into force, and $Q(s)$ converts force back into quadrature
fields. $f$ are spurious forces that act on the interferometer
mirrors.}
\end{figure}

To stabilize this resonance, we apply a velocity damping force via
a feedback control system; a schematic for the control system is
shown in Fig.~\ref{fig:feedback_loop}. Ordinarily, we are
interested in the (squeezed) output field $b$ that exits the
ponderomotive interferometer (IFO), but we need to detect a small
fraction of $b$ to generate a control signal for damping the
unstable resonance. We, therefore, insert a beamsplitter (BS) at
the IFO output and use the field $u = \sqrt{R}\,b$ ($R \ll 1$) in
a feedback loop. The quadrature field $u$ is converted into a
force by the transfer function $F(s)$ and $Q(s)$ converts force to
quadrature fields. The velocity damping term is included in
$F(s)$. $Q(s)$ contains the force-to-displacement transfer
function $P(s)$ [see Eqn.~(\ref{eq:pend})], as well as the
input-output relation that converts displacement to quadrature
field [see e.g. Eqns.~(63) and (64) of Ref.~\cite{code_paper}].
The majority of the squeezed field, $y = \sqrt{T}\,b$, is
preserved as a squeeze source. Vacuum noise fields $n_0$, $n_{c1}$
and $n_{c2}$ enter the open ports of the beamsplitter, and must be
accounted for in the total noise budget.

Defining the open-loop gain of the feedback system as

\begin{equation}
G(s) = - \sqrt{R}\,F(s)\,\,Q_{\zeta}(s)\,,
\end{equation}
the squeezed output field $\textbf{y}$ is given by

\begin{eqnarray}
y_{\zeta} &=& \frac{\sqrt{T} {\mathcal M}_{\zeta} \cdot
\textbf{a}}{1 + G(s)} + \frac{Q_{\zeta} \sqrt{T}}{1 + G(s)}\,
\textbf{f} \nonumber
\\ \label{eq:yzeta} & & + \frac{T\,G(s)}{1 + G(s)}
\frac{1}{\sqrt{R}}\,\left(n_{c1}\right)_{\zeta} - \sqrt{R}
\left(n_{c1}\right)_{\zeta}\,,
\end{eqnarray}
where ${\mathcal M}$ is a matrix operator that converts the input
field $\textbf a$ to the output $\textbf{b}$, $Q$ converts forces
into quadrature fields, and the subscript $\zeta$ denotes the
projection on the quadrature to be measured. Eqn.~(\ref{eq:yzeta})
warrants some discussion. The first term contains the squeezed
output due to the input field $\textbf{a}$. In order to realize
the squeezing without the influence of the control system, it is
necessary to make $G(s)$ as small as possible in the band where
squeezing is to be measured. Similarly, when $G(s) \gg 1$, the
last term dominates and $R$ should be kept small to couple as
little of the vacuum noise $\left(n_{c1}\right)_{\zeta}$ to the
output $y_{\zeta}$.\footnote{We do not combine the last two terms
containing $\left(n_{c1}\right)_{\zeta}$ because we will assume
that those two terms are uncorrelated. This is not true, but will
at worst give an underestimation by a factor of $2$ of the noise,
and for the cases when $\left|G\right|\ll 1$, the error is much
smaller.} Finally, to stabilize the optomechanical resonance, we
need to introduce a damping term to $P(s)$ [implicitly included in
$G(s)$]. We propose a filter transfer function that is equivalent
to applying a velocity damping:
\begin{equation}
F = \frac{-s\gamma_{d}}{\sqrt{R}}
\end{equation}
where $\gamma_{d}$ is a damping constant chosen to stabilize the
system. The open loop gain then becomes
\begin{equation}
G(s) = s\,\gamma_{d}\left[s^{2} + \frac{\Theta^{2}
\gamma^{2}}{\left(\gamma + s\right)^2+ \delta^2}\right]^{-1} \,,
\end{equation}

In addition to stabilizing the optomechanical resonance, we must
minimize the additional noise due to vacuum fluctuations that are
introduced by the new beamsplitter. We consider only the newly
i8ntroduced vacuum noise that is detected by the feedback
detector, which is then fed back onto the position of the pendulum
and thereby enters the signal detected by the squeeze detector. We
neglect the correlations between these vacuum fluctuations that
enter directly at the beamsplitter with those that enter through
the feedback loop. This is a valid assumption for frequencies at
which $\left|G(s)\right| \ll 1$, which is the case in our
measurement band. Assuming that the feedback detection is
shot-noise-limited, then the power spectral density of the
additional noise, relative to shot noise, is
\begin{equation}
S_{n} \leq \sqrt{\left |\frac{G(s)}{1 + G(s)} \right
|\frac{1}{\sqrt{R}} + \sqrt{R}}
\end{equation}
[see the last two terms in Eqn.~(\ref{eq:yzeta}), with $R \ll 1$
so that $\sqrt{1-R} \approx 1$]. Choice of 3 to 10\% for the
nominal value of $R$ gives acceptable levels of loss for the
squeezed output beam, while allowing for feedback. We note that
for the case $\left| G(s) \right | \gtrsim 1$, that these
expressions are not valid, and a detailed calculation of the
correlations must be done. The correlation between the last two
terms in Eqn.~(\ref{eq:yzeta}) depends on the quadrature being
measured to do the feedback; we assume the worst case scenario for
the noise, namely that the two terms add in amplitude.

In order to keep the coupling of vacuum noise $n_{c1}$ into
$y_{\zeta}$ at a minimum, we must make the loop gain $G(s)$ as
small as possible at frequencies within the squeezing measurement
band (about 100 Hz to 1 kHz), while still having sufficient gain
at the optomechanical spring resonance frequency (typically 5
kHz). We achieve this by including a sharp high-pass filter in
$F(s)$, typically an elliptic filter with high-pass corner
frequency at a several 100 Hz to preserve phase margin at the
optical spring resonance. The resulting contribution to the
overall noise budget is show as the curve labeled ``Control
noise'' in Fig.~\ref{fig:fig1}, where we set $\gamma_{{\rm d}} = 7
\times 10^{4} {\rm ~s}^{-1}$, $R = 3\%$, and a fourth-order
elliptic high-pass filter with cut-off frequency at 800~Hz. A
detailed analysis of the control system can be found in
Ref.~\cite{control_doc}.

\subsection{Laser Noise}
\label{sect:lnoise}

\begin{figure}[t]
\includegraphics[height=7cm]{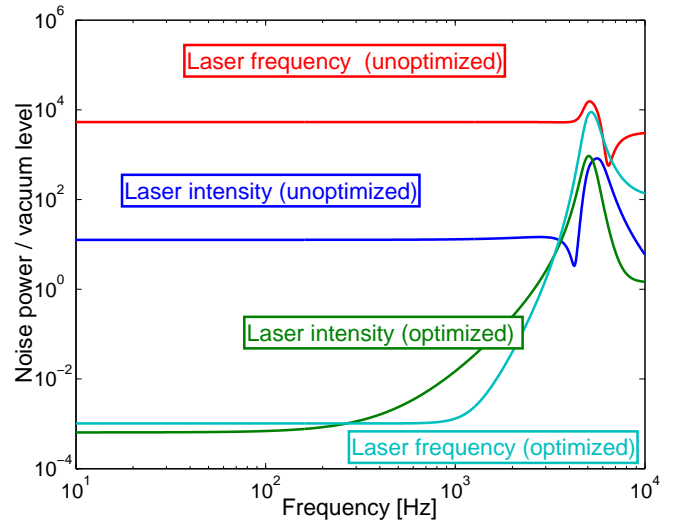}
\caption[Figure] {\label{fig:lnoise} The coupling of laser noise
to the antisymmetric port is shown for the unoptimized and
optimized cases. The optimized case includes a Michelson detuning
and intentional loss in one of the arms.}
\end{figure}

Laser intensity and frequency noise couple to the output port of
the interferometer through imperfections and mismatch in the
optical parameters of the interferometer. Analytic calculation of
such noise couplings were carried out in Ref.~\cite{code_paper}.
The calculations lead to complex formulae that, in our opinion, do
not provide much insight into the couplings, except the following
qualitative features. For frequencies much below $\Theta$ and
$\gamma$, and up to leading order of $\Theta L / c$, $\gamma L /
c$ and $\delta L /c$, phase and amplitude noises both emerge in
single quadratures (as a result, there exist a phase-noise-free
quadrature, and an amplitude-noise-free quadrature.) The phase
noise does not drive mirror motion, and emerges at the output at
an orthogonal quadrature to the carrier leaking out from that port
(i.e., the carrier coincides with the phase-noise-free
quadrature). The amplitude noise, on the other hand, drives mirror
motion, and emerges in a quadrature neither along nor orthogonal
to the carrier. Different types of mismatches direct laser
amplitude and phase noises into different output quadratures. Up
to linear order in mismatch, the output phase (amplitude) noise
can be expressed in the quadrature representation as a sum of
quadrature vectors, each arising from one type of mismatch.

\begin{figure*}[t]
\includegraphics[width=18cm]{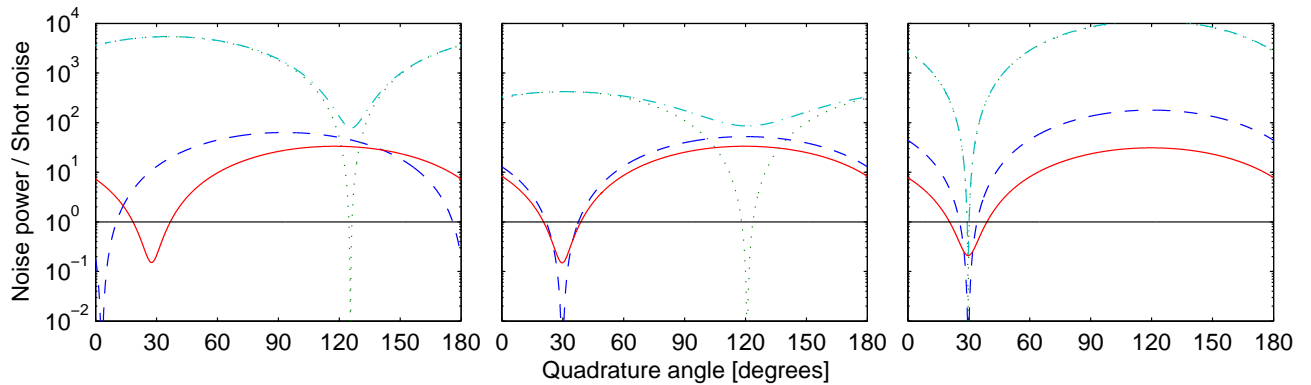}
\caption[Figure] {\label{fig:quad3}
The coupling of laser and
antisymmetric port noise to the output as a function of the
homodyne measurement quadrature for the unoptimized case. The red
curves represent the quantum optical noise, the blue curves
represent laser intensity noise, the green curves represent laser
phase noise and the cyan curves represent the total noise. In (a),
the minimal noise quadratures for the different noise sources are
not the same. In (b), the minimal noise quadratures for the laser
intensity noise and the vacuum fluctuations are now the same. For
this case, only a Michelson detuning has been added. In (c), the
minimal noise quadratures for the laser intensity noise, laser
frequency noise and the vacuum fluctuations are now the same. For
this case, a Michelson detuning and a controlled loss in one arm
(between the beamsplitter and the input test mass) were used. }
\end{figure*}

In full numerical results, we did not observe phase-noise-free and
amplitude-noise-free quadratures, but instead found output
quadratures in which contributions from one of the two laser
noises has a rather deep minimum. The minimum-phase-noise and
minimum-amplitude-noise quadratures do not generically agree with
each other, nor do they generically agree with the
minimum-quantum-noise quadrature. However, we have discovered that
it is possible, by intentionally introducing controlled
mismatches, to modify the quadrature dependence of both of the
output laser noises in such a way that both the
minimum-phase-noise and minimum-amplitude-noise quadratures align
with the minimum-quantum-noise quadrature. Such a procedure
greatly reduces the importance of the laser noise, {\it as long as
the noise in the minimum noise quadrature is concerned}. This is
shown in Figs.~\ref{fig:lnoise} and~\ref{fig:quad3}, using our
fiducial parameters in Table~\ref{table:param}.

Let us describe the optimization procedure in more detail. Through
the numerical simulation~\cite{code_paper}, we determine that the
noise quadratures may be optimized through two steps, as shown in
Fig.~\ref{fig:quad3}. The first step is to detune the Michelson
from the dark fringe. The optimal position for the Michelson
detuning is that which aligns the minimum-amplitude-noise
quadrature to the minimum-quantum-noise quadrature. The second
step is to introduce an intentional loss into one arm of the
Michelson, placed artificially between the beamsplitter and one of
the arm-cavity mirrors, such that both minimum-amplitude-noise and
minimum-phase-noise quadratures would align with the
minimum-quantum-noise quadrature. Interestingly, since the
minimum-phase-noise quadrature coincides with the carrier
quadrature leaking out from the output port, the resulting
squeezed output light is amplitude squeezed.

As it turns out, the required artificial loss can be quite large;
for our fiducial parameters in Table~\ref{table:parameters}, the
optimal loss is approximately 10\%. Such a large loss will
noticeably limit the amount of squeezing that may be detected, but
the reduction in the laser noise is necessary to measure any
squeezing at all. As shown in Figs.~\ref{fig:lnoise}
and~\ref{fig:quad3}, the laser amplitude noise (as measured in the
squeezed quadrature) is reduced by more than 40 dB and the laser
frequency noise by more than 60 dB in this process --- both of
them now are far below the quantum noise level.

It is difficult to predict exactly the mismatches that will be
present in the physical experiment. Rather than making a priori
predictions for the intentional mismatch needed to optimize the
noise couplings, we plan to perform this optimization empirically.
We estimate that the ability to control the loss at the level of
$0.1\%$ and the detuning at the level of $10^{-4}\delta_\gamma$ is
sufficient for the optimization.

Although we have greatly reduced the laser noise in the ideal
quadrature, we have not reduced its overall magnitude. This
presents a limitation because we must control the quadrature
measurement angle to be precisely at the ideal quadrature. Small
fluctuations in this measurement angle will couple noise in from
the orthogonal quadrature, where the noise is much larger. This is
evident from the sharp features in Fig.~\ref{fig:quad3}, which
shows that the margin for error in the measurement quadrature is
quite narrow due to the laser frequency noise.

\subsection{Quantum noise and losses}
The quantum noise, due to output port vacuum fluctuations and
optical losses, are also calculated using the noise simulation
code~\cite{code_paper}. Considering only the noise that enters
through the output port, and neglecting other noise sources,
including optical losses, the vacuum field is squeezed by ~$17$ dB
inside the interferometer.

Next, we include optical losses at the levels given in
Table~\ref{table:parameters}. In particular, our simulation code
has automatically taken into account intracavity losses, losses
due to transmission through the IMs, losses of the beamsplitter,
losses into the common mode due to mismatches, and artificial
losses introduced to mitigate laser noise in the detected
quadrature. These together lead to a noise spectrum at the level
of $\sim 7$ dB below shot noise (see Fig.~\ref{fig:fig1}). We
expect this to be the limit to measurable squeezing in most of our
frequency band.

\subsection{Summary of design considerations}
Considerations of the detailed parameters of the experiment is a
sequence of trade-offs between achieving high levels of squeezing
and keeping the noise couplings to a minimum. In
Table~\ref{table:choices} we summarize the highly intertwined and
often conflicting considerations that informed the design in the
preceding sections.

 \begin{table*}[t]
 \begin{tabular}{lllllll}
 \hline Parameter & \hspace{0.15cm} Advantages of large value & \hspace{0.15cm} Advantages of
small value\\
 \hline \hline
ETM mass & Ease of construction & Large optical spring frequency\\
 & Ability to sense and actuate
motion & \vspace{0.05cm}\\
\hline ITM transmission & Large optical spring frequency & Reduce
optical spring instability \\ (Cavity finesse) & & Reduce
effective intracavity
losses\\ & & Higher circulating power could damage mirrors \vspace{0.05cm}\\
\hline ITM mass & Ease of construction & Increase optical spring
frequency \footnote{Making the ITM mass the same as that of the
ETM, for example, would increase
the optical spring resonance frequency by $\sqrt{2}$.} \\
& Work with existing sizes and solutions & \vspace{0.05cm}\\
\hline Input power & Large optical spring frequency & Use
available
lasers\\ & & Stay below damage threshold of cavity mirrors \vspace{0.05cm}\\
\hline Detuning or $\delta_\gamma$ & Optimize $\delta_\gamma = 1/\sqrt{3}$ for
largest & Use smaller $\delta_\gamma$ to increase squeezing level
\vspace{-0.1cm} \\ & squeezing
bandwidth & \vspace{0.05cm}\\
\hline Spot size on ETM & Reduce coating thermal noise &
Reduce angular instability of cavity \vspace{0.05cm} \\
\hline Cavity length & For fixed beam size on mirror surfaces, &
Reduce instability of optomechanical resonance \vspace{-0.1cm} \\
& longer length increases
suppression of & \vspace{-0.1cm} \\ & higher order spatial modes & \\
& Larger mirror radii of curvature easier & \vspace{-0.1cm} \\ & to manufacture &  \vspace{0.05cm}\\
\hline
\end{tabular}
\caption{Design considerations for select interferometer
parameters. Here we tabulate some of the competing effects that
led us to the choice of parameters listed in
Table~\ref{table:parameters} and discussed in Sections
\ref{sect:design} and \ref{sect:noise}. }\label{table:choices}
\end{table*}

\section{Summary and Conclusions}
\label{sect:conclusions}

We have presented a design for an interferometer with movable
light mirror oscillators, such that the light (and vacuum) fields
circulating in the interferometer are squeezed due to the coupling
of radiation pressure and motion of the mirrors. We show that even
in the presence of reasonable, experimentally realizable optical
losses (at the level of $10^{-5}$ per bounce per optic), thermal
noise (associated with oscillators with intrinsic loss factors of
order $10^{-7}$), and classical laser noise (relative intensity
noise at the level of $10^{-8}$ and frequency noise $10^{-4} {\rm
~Hz/}\sqrt{{\rm Hz}}$), significant levels of squeezing can be
extracted from such a device. Specifically, we find that as much
as 7~dB of squeezing at 100~Hz is possible, provided great care is
exerted to measure the quadrature where the laser noise coupling
to the output is minimized, as shown in Fig.~\ref{fig:fig1}. We note
that the squeezed state produced by this device will be far from a
minimum uncertainty state (the noise in the anti-squeezed quadrature
relative to the squeezed quadrature is much greater than required by
Heisenberg's Uncertainty Principle). This will place requirements on the
stability requirements for any device to which the state is applied.

Two aspects of the design require great care: the optical
performance of the high finesse, detuned arm cavities (described
in Section~\ref{sect:optical_design}); and the mechanical design
of the suspended 1~gram mirror oscillators, where thermal noise
must be kept at low, and pitch, yaw and longitudinal degrees of
freedom must be controllable by application of external forces
outside the measurement band (described in
Section~\ref{sect:mech_design}).

This is, to our knowledge, the first viable design for extracting
the squeezing generated by radiation-pressure-induced rigidity in
an interferometer, and construction of this experiment is underway
at our laboratory.

\acknowledgments We thank our colleagues in the LIGO Scientific
Collaboration, especially Vladimir Braginsky, Keisuke Goda,
Eugeniy Mikhailov and Gregg Harry, for stimulating discussions. We
gratefully acknowledge support from National Science Foundation
grants PHY-0107417, PHY-0300345, PHY-0099568 (for Y.C.), and
PHY-0353775 (for F.K. and S.V.). Y.C.'s research was also
supported by the David and Barbara Groce Fund at the San Diego
Foundation, as well as the Alexander von Humboldt Foundation's
Sofja Kovalevskaja Award (funded by
the German Federal Ministry of Education and Research). F.K. and
S.V.'s research was also supported by the Russian Agency of
Industry and Science contracts \#40.02.1.1.1.1137 and
\#40.700.12.0086, and by Russian Foundation of Basic Researches
grant \#03-02-16975-a.

\end{document}